\documentstyle[11pt,aaspp4,psfig]{article} 

\slugcomment{Accepted for publication in {\it Icarus}, July 2, 2000}

\lefthead{Reach et al.}
\righthead{Encke Dust Trail} 

\begin{document}

\title{The formation of Encke meteoroids and dust trail
\footnote{Based on observations with ISO, an ESA project with
instruments funded by ESA Member States (especially the PI countries:
France, Germany, the Netherlands and the United Kingdom) with the
participation of ISAS and NASA.}}

\author{William T. Reach}
\affil{Infrared Processing and Analysis Center,
California Institute of Technology,
Pasadena, CA 91125}

\author{Mark V. Sykes}
\affil{Steward Observatory, University of
Arizona, Tucson, AZ 85721}

\author{David Lien}
\affil{Washington and Jefferson College,
Washington, PA}

\author{John K. Davies}
\affil{Joint Astronomy Center, 600 N. A'ohoku
Place, Hilo, HI 96720}

\def\etal{ et al. }
\def\ISO{ {\it ISO} }
\def\arcsec{\hbox{$^{\prime\prime}$}}
\def\kms{ {km~s$^{-1}$}}
\def\simgt{\mathrel{\mathpalette\oversim>}}

\begin{abstract}

\def\textabs{
We observed comet 2P/Encke with the Infrared Space Observatory ISOCAM on July 14,
1997 from a particularly favorable viewing geometry above
the comet's orbital plane and at a distance of 0.25 AU. 
A structured coma was observed, along with a long, straight dust trail. 
For the first time, we are able to observe the path of particles
as they evolve from the nucleus to the trail. The particles that produce the
infrared coma are large, with a radiation to
gravitational force ratio beta<0.001
(corresponding to >mm-sized particles).
The dust trail follows the orbit of the comet across our
image, with a central core that is 20,000 km wide,
composed of particles with beta<1e-5 (size $\sim 5$ cm)
from previous apparitions.
The abundant large particles near the comet pose a 
significant hazard to spacecraft. 
There is no evidence of a classical cometary dust tail due to small
particles with beta>0.001, in marked contrast to other comets like 
P/Halley or C/Hale-Bopp.
The structure of the coma requires anisotropic emission and that
the spin axis of the nucleus 
to be nearly parallel to the orbital plane, resulting in strong seasonal
variations of the particle emission.
While most of the infrared coma emission is due to dust
produced during the 1997 apparition, the core of the dust trail
requires emissions from previous apparitions. 
The total mass lost during the 1997 apparition is estimated to be
 2-6e13 g.
Comparing to the gas mass loss from ultraviolet observations, the
dust-to-gas mass ratio is 10-30, much higher than has ever been
suggested from visual light observations.
Using the recently-measured nuclear diameter,
we find that Encke can only last 3000-10,000 rhoN yr (where rhoN 
is the nuclear density in g/cc) at its present mass loss rate.
}

We observed comet 2P/Encke with the {\it Infrared Space Observatory} ISOCAM on July 14,
1997 from a particularly favorable viewing geometry above
the comet's orbital plane and at a distance of 0.25 AU. 
A structured coma was observed, along with a long, straight dust trail. 
For the first time, we are able to observe the path of particles
as they evolve from the nucleus to the trail. The particles that produce the
infrared coma are large, with a radiation to
gravitational force ratio $\beta<10^{-3}$
(corresponding to $>$mm-sized particles).
The dust trail follows the orbit of the comet across our
image, with a central core that is $2\times 10^4$~km wide,
composed of particles with $\beta<10^{-5}$ (size $\sim 5$ cm)
from previous apparitions.
The abundant large particles near the comet pose a 
significant hazard to spacecraft. 
There is no evidence of a classical cometary dust tail due to small
particles with $\beta>10^{-3}$, in marked contrast to other comets like 
P/Halley or C/Hale-Bopp.
The structure of the coma requires anisotropic emission and that
the spin axis of the nucleus 
to be nearly parallel to the orbital plane, resulting in strong seasonal
variations of the particle emission.
While most of the infrared coma emission is due to dust
produced during the 1997 apparition, the core of the dust trail
requires emissions from previous apparitions. 
The total mass lost during the 1997 apparition is estimated to be
 2--6$\times 10^{13}$ g.
Comparing to the gas mass loss from ultraviolet observations, the
dust-to-gas mass ratio is 10--30, much higher than has ever been
suggested from visual light observations.
Using the recently-measured nuclear diameter,
we find that Encke can only last 3000-10,000 $\rho_N$ yr (where $\rho_N$ 
is the nuclear density in g~cm$^{-3}$) at its present mass loss rate.

\end{abstract}

\section{Introduction}

The discovery of cometary dust trails by the {\it Infrared Astronomical 
Satellite} (Davies {\it et al.} 1984, Sykes {\it et al.} 1986) revealed
that comets emit far more dust than expected based on visual
wavelength observations (Sykes and Walker 1992). The trails consist of 
large, dark particles (in the millimeter to centimeter size range),
and they were
found to represent the principal mass loss mechanism for the comets with which
they were associated.  
From the {\it IRAS} observations, it was inferred that dust trails were a 
phenomenon common to short-period comets in general, and that the `dirty
snowball' of Whipple (1950) was more accurately a `frozen mudball'
(Sykes 1993).
Even the classical `gassy' comet P/Encke was observed to have a trail, 
and it was shown that the dust to gas mass ratio of the material ejected from 
the comet was 3.5, as opposed to the canonical 0.1 to 1
(Sykes and Walker 1992). 

We will use the terms `meteoroid' and `dust particle' synonymously in this
paper, but it should be evident from our results 
that the `dust particles' we are talking
about are larger than those often referred to by observers of other
comets. In fact, the `dust particles' are the same size as the particles
that give rise to meteors when they enter the Earth's atmosphere
(Ceplecha 1998).
Thus when we describe the `Encke dust trail,' we are describing the
origin of meteoroids from comet Encke. Comet Encke is the parent body
of the Taurid meteor stream complex, so we infer that the `Encke dust trail'
is the set of recently-produced particles that would become a meteor
stream if their orbit crossed the Earth's orbit (as the Taurid meteoroids'
orbits now do).
Meteoroids and dust particles are ejected from comets due to pressure from
sublimating ices, and after being accelerated to some terminal velocity ($v_{ej}$),
the principal forces on optically large particles are solar gravity and 
radiation pressure (parameterized by the force ratio 
$\beta=F_{rad}/F_{grav}$).
The width of the dust trails perpendicular to the orbital plane,
observed by {\it IRAS}, constrained $v_{ej}$ and $\beta$.
For P/Encke, the trail spanned a $90^\circ$ of mean anomaly,
or about 2 AU of physical length.
The length of a dust trail ahead of and behind the comet
allows an estimate of the ages of the trails, which are typically
many orbital periods of the comets
(Sykes {\it et al.} 1990, Sykes and Walker 1992). 
However, the relatively low angular resolution of {\it IRAS}
limited the information that could be easily extracted from the data.

Frequent opportunities exist to study comets from ground-based telescopes,
but low albedos and optical depths of $\tau\sim 5\times10^{-9}$ make trails
extremely difficult to observe at visible wavelengths. In the thermal infrared
they are only available from space-based platforms. 
The apparition of P/Encke in 1997 provided a special opportunity 
to study the dynamics and distributions of various size ranges of particles
ejected from the nucleus, because we could see the comet from a favorable
viewing geometry above its orbital plane.
The favorable geometry, increased sensitivity, higher angular resolution of
the {\it Infrared Space Observatory} camera relative to {\it IRAS}, 
and proximity of Encke
to the Earth, made the 1997 apparition of Encke a prime opportunity
to advance our understanding of the nature and origin of cometary dust.

\section{{\it ISO} Observations}

We observed comet Encke using the {\it Infrared Space Observatory}
({\it ISO}; Kessler {\it et al.} 1996)
on July 14, 1997, after Encke had passed perihelion on May 23, 1997. 
At the time of observation Encke was at a distance
$\Delta=0.25$ AU from the Earth and $R=1.15$ AU from the Sun. 
It was observed from a vantage point well above the 
orbital plane, with the angle between the line of sight
and the orbital plane being $35^\circ$.
The phase angle (Sun-Encke-Earth) was $53^\circ$ for this observation.
Figure~\ref{enckeday} shows the locations of Encke and Earth in their
orbits at the time of observation.
For reference,
Encke's orbit has semimajor axis 2.21 AU, eccentricity 0.85,
and inclination $11.9^\circ$.

\centerline{\psfig{figure=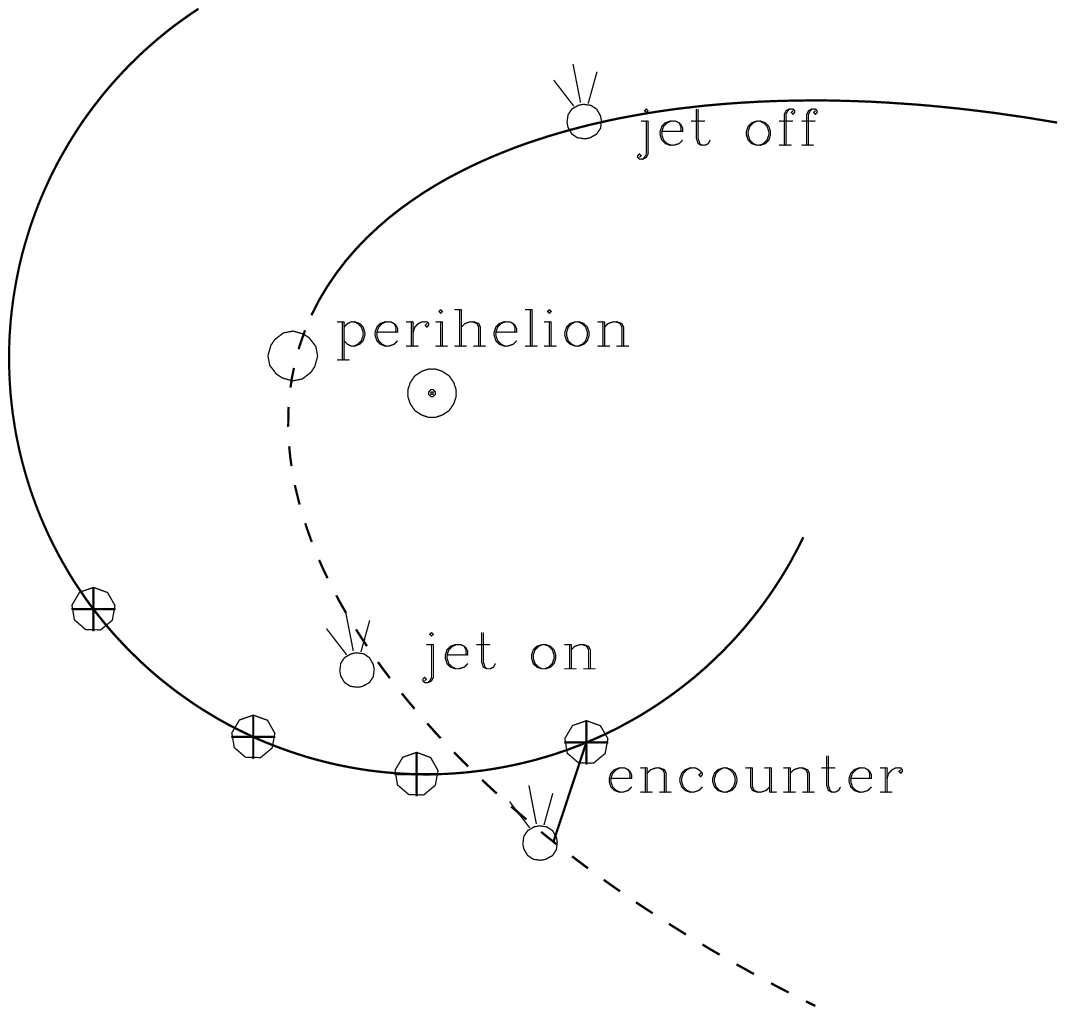}} 
\figcaption{\it Locations of Encke and Earth in their orbits for 4 of the
events discussed in this paper. Encke's orbit is shown as a dashed line
where it is below the Earth's. Our {\it ISO} observation,  
indicated by the
line connecting Encke and Earth, occurred on July 14, 
51 days after perihlion.
In the model presented in this paper, the dust production occurs
primarily from a single jet. 
The locations of Encke and Earth when the jet turns off (about 30 days before
perihelion) and on (about 25 days after perihelion) are indicated with
symbols on their respective orbits. The symbol for Encke shows the
approximate position angle of the jet with three lines pointing in the
direction of particle ejection.
\label{enckeday}}

For our observations, we
used the mid-infrared camera ISOCAM (Cesarsky {\it et al.} 1996) with the
$6^{\prime\prime}$ pixel-field-of-view lens and the widest filter (LW10, nominal
wavelength 11.5 $\mu$m, range 8--15 $\mu$m). ISOCAM consists of a
$32\times 32$ detector array, so in this configuration the 
instantaneous field of
view is $3.2^\prime\times 3.2^\prime$.
The ideal diffraction limit is $5^{\prime\prime}$, approximately the same
as the pixel size, so a typical point source will have a full width
at half maximum of less than 2 pixels.

Owing to its rapid motion ($7^\prime$/hr) and large angular size ($>
20^\prime$), comet Encke was a challenging target for {\it ISO}
observations. We devised a strategy that allowed us to cover a large
area with high sensitivity, without allowing the comet to smear or
the stars to trail. To do this, we made a series of pointed
observations, taking 11 exposures of 2.1 sec duration at each
position. Encke's motion during the 23 sec spent at each position is
less than half a pixel, so each set of 11 frames can be coadded
without smearing the comet. To build a mosaic large enough to show
extended features, we pointed the telescope in a $21\times 11$
raster, with step sizes of $90^{\prime\prime}\times
76^{\prime\prime}$, oriented perpendicular to the comet's motion. The
raster leg length was chosen to be long enough to cover the desired
area, yet fast enough that the comet could not move more than half of
the ISOCAM field of view during a raster leg. This ensures that we
saw the comet on two raster legs. The spacing between raster legs was
chosen to be equal to the motion of the comet during an individual
raster leg. This is the optimal strategy for covering the widest
possible area in comet-centered coordinates, while maintaining enough
overlap between frames to allow accurate removal of the
pixel-to-pixel gain variations.

Because the {\it ISO} operation system did not allow raster
observations of moving targets, we negotiated with the {\it ISO} team
for an observing date, close to Encke's passage near the Earth, that
fit within the {\it ISO} schedule and viewing constraints. For each
of the possible observing dates, we calculated the position and
orientation of motion of Encke for that observing date, and we
specified a window for scheduling. The observations spanned the period
between 00:07 and 02:03 UT.  If the observation had been
scheduled earlier or later, part or all of the desired field would
have been missed.
(Unfortunately, the other comet in this observing program, 81P/Wild 2, 
was missed entirely.)
Because the Encke observation took place exactly at
the time for which we calculated the ephemeris, Encke is centered in
the final map. In celestial coordinates, the map covers 
$15.7^\prime \times 33^\prime$, and in cometocentric coordinates,
the map covers $30^\prime \times 33^\prime$.

The set of 2541 exposures were reduced as follows. First, cosmic rays
were identified using a running median on the time history of the
brightness seen by each pixel. For each cosmic ray, the affected
pixel was flagged for 4 exposures beginning with the cosmic ray
event; this helps remove pixels affected by transient gain variations
induced by the cosmic ray. Then the dark current was subtracted from
each exposure using a library image of the dark current from the
routine calibration observations. Then the time history of each pixel
was corrected for the transient gain response, by deconvolving the
the time series from the response curve, which consists 
of a fast response to 60\% of a brightness change followed by a slower response
to the remaining 40\% with a time constant $\tau=320/I$ sec for a
brightness $I$ MJy~sr$^{-1}$; the transient corrections make
only small changes to the image because the comet and zodiacal light
were very bright for this observation (cf. Coulais and Abergel 1999).
The 10 exposures
at each position for which the telescope was not slewing
were coadded to give 242 images. The pixel-to-pixel
gain response and the optical vignetting were corrected by dividing
each frame by a flat-field image, constructed from
the median of all images excluding images taken
within $10^\prime$ of the comet nucleus. 

\centerline{\psfig{figure=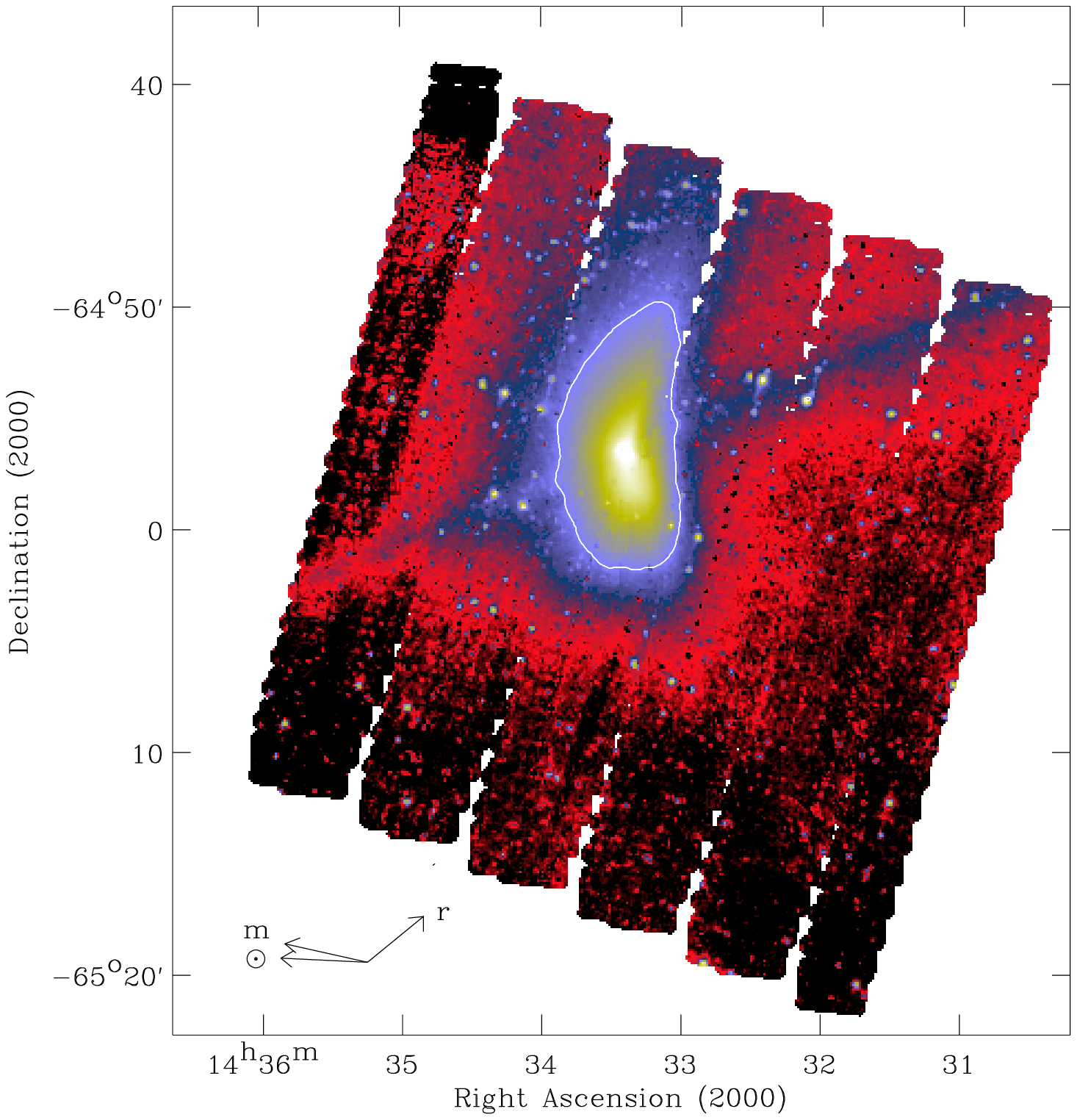,width=8truein,angle=90}} 

\figcaption{\it Image of comet Encke in the comet rest frame, made from observations
by the {\it Infrared Space Observatory} in 1997. This image is a
mosaic of 231 individual pointings with an instantaneous field of
view of $3^\prime \times 3^\prime$. Each of the individual pointings
was shifted (before coaddition into the final mosaic) in reflex with
the comet's motion. Stars appear as multiple point sources, one for
each raster leg that crosses the star's position, spread out along a
straight line in the direction of the comet's motion. 
The color table goes from black (no emission) to red (faint emission,
1 MJy~sr$^{-1}$) to blue (moderate emission, 4 MJy~sr$^{-1}$) to 
greed (bright emission, 15 MJy~sr$^{-1}$) to 
white (very bright emission, 100 MJy~sr$^{-1}$). 
A contour is shown at 10 MJy~sr$^{-1}$;
the color table switches at this contour from a linear table for faint to
moderate emission to a logarithmic table for bright emission.
The direction to the Sun is shown as a vector labeled `$\odot$';
the direction of the comet's motion is shown as a vector labeled `m';
and the direction of the rotational pole is shown as a vector
labeled 'r'.
The dust trail
is clearly visible in this image, stretching diagonally from the
lower left to upper right. \label{isoshift}}

Figure~\ref{isoshift} shows a mosaic of the ISOCAM images
in the rest frame of comet Encke.
Each of the 242 images was shifted to match
the comet's motion relative to its position halfway through the
observation. Therefore, this image simulates a `snapshot' of the
comet on July 14, 1997 at 01:00 UT. In Fig.~\ref{isoshift}, stars
appear as a pair of point sources spread along the direction of the
comet's motion (roughly right to left). The spacing between the star
images depends on the interval between crossings of the star during
the raster (which was performed with legs of alternating direction).

\centerline{\psfig{figure=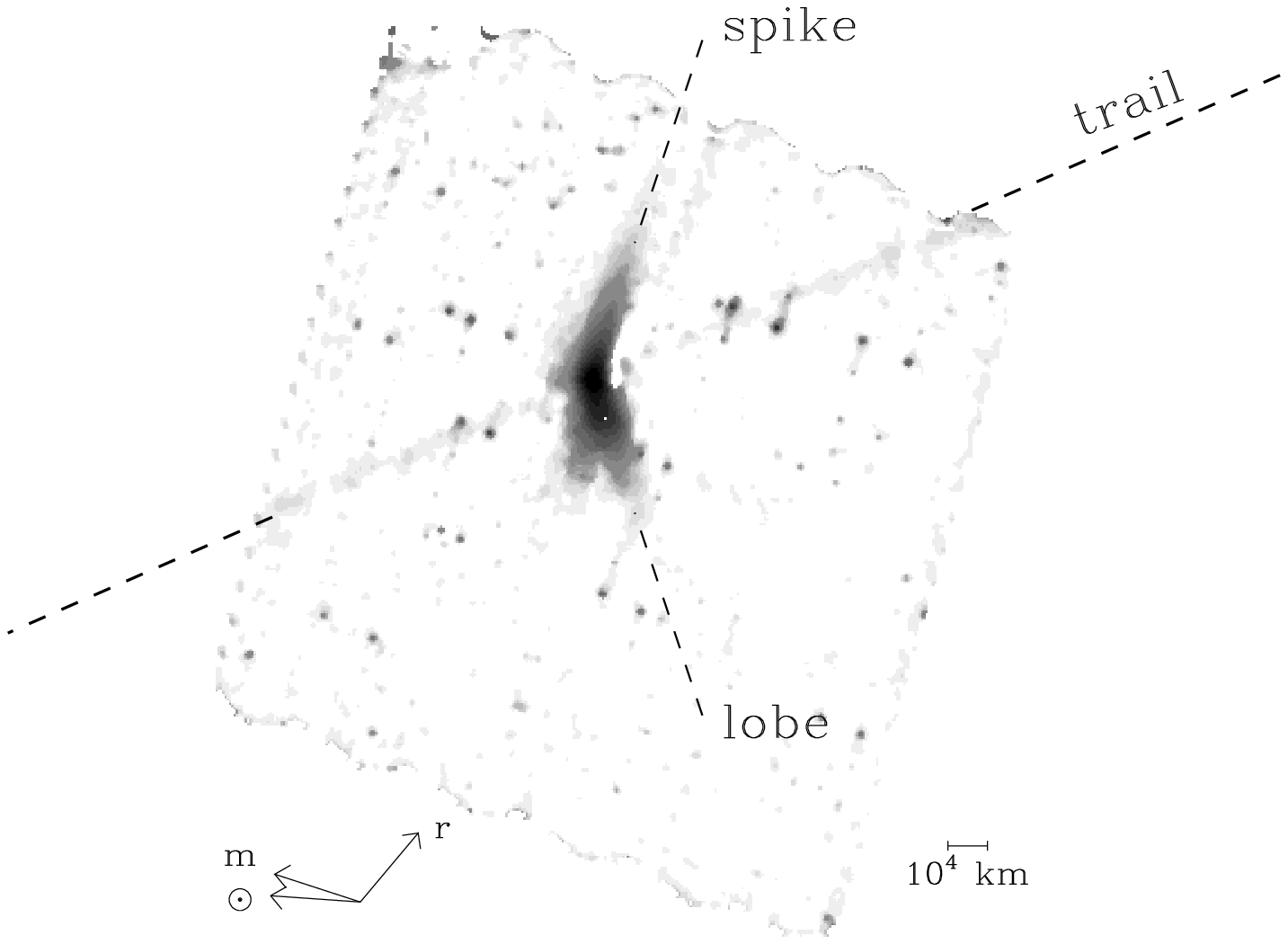}} 

\figcaption{\it Median-filtered version of the ISOCAM image of P/Encke.
Each pixel in this image is the median in a $3\times 3$ pixel box
of the original image minus the median in a $31\times 31$ pixel box.
 The size and orientation
are the same as in Fig.~\ref{syndynes}. The `NW spike,'
the `louthern lobe,' and the debris `trail' are labeled.
\label{spikefig}}

Several structures can be distinguished in our
ISOCAM observation of comet Encke. 
Figure~\ref{spikefig} shows an unsharp-masked version of the ISOCAM image,
with several features labeled.
The structures include a coma, consisting of a lobe
extending to the south and a spike extending to the northwest,
and a thin, linear trail that stretches across the
image, running directly through the nucleus
from the upper right (following the comet) to the lower left 
(leading the comet).
The trail is brighter following the comet than leading, and
there is a steeper gradient of brightness with distance leading the
comet than following the comet. 
The vertical profile of the trail is also somewhat different for
the portions of the trail leading and following the comet.
Figure ~\ref{trailprof_encke} shows 
two perpendicular brightness profiles of the dust trail. 
The two cuts are at angular separations of $\pm 12^\prime$ from the 
nucleus, corresponding to a separation from the nucleus by an
orbital mean anomaly of $\pm 0.014^\circ$ and a physical
distance of approximately $\pm 1.5\times 10^5$ km.
The profile
leading the comet is consistent with a single Gaussian component, while
the profile following the comet is very different.
In addition to a nearly Gaussian core (with width similar to that of
the trail ahead of the comet), there is a wider, underlying component, as
well as significant emission from the {\it tail}, which curves over the
trail behind the comet. For the purpose of this paper, we fit the
trail as the sum of two Gaussian components, labeled `core' and `skirt'
in Fig.~\ref{trailprof_encke}. 
Table ~\ref{proftab} shows the parameters
of Gaussian fits to the brightness profiles.
The optical depth was calculated for the peak brightness through the
trail: $\tau=I_\nu/B_\nu(T)$, where $B_\nu$ is the Planck function.
We assume a temperature $T=270$~K,
based on the time of the ISOPHOT spectrophotometry of the 
coma (Lisse et al. 2000). For reference, $B_\nu(270{\rm~K})=2.6\times 10^8$
MJy~sr$^{-1}$ at 11.7 $\mu$m (the center of the response curve
for the LW10 filter).
As discussed below, the particles in the
dust trail are likely to be larger than those in the coma and tail.
But even the coma and tail particles appear to be large enough that they
emit as greybodies: their temperature is only 4\% warmer than expected for
rapidly-rotating grains larger than 10 $\mu$m radius, and they have
a grey emissivity even out to 100 $\mu$m wavelength (Lisse et al. 2000).
Very large grains may preserve a day-night temperature difference 
across their surface, which could elevate the temperature of the trail
particles somewhat. Sykes and Walker (1992) determined that Encke
trail particles had temperatures $< 6$\% in excess of a greybody.
The temperature excess is likely
to be less than 15 K, meaning our optical depths are unlikely
to be wrong by more than 25\%.

\centerline{\psfig{figure=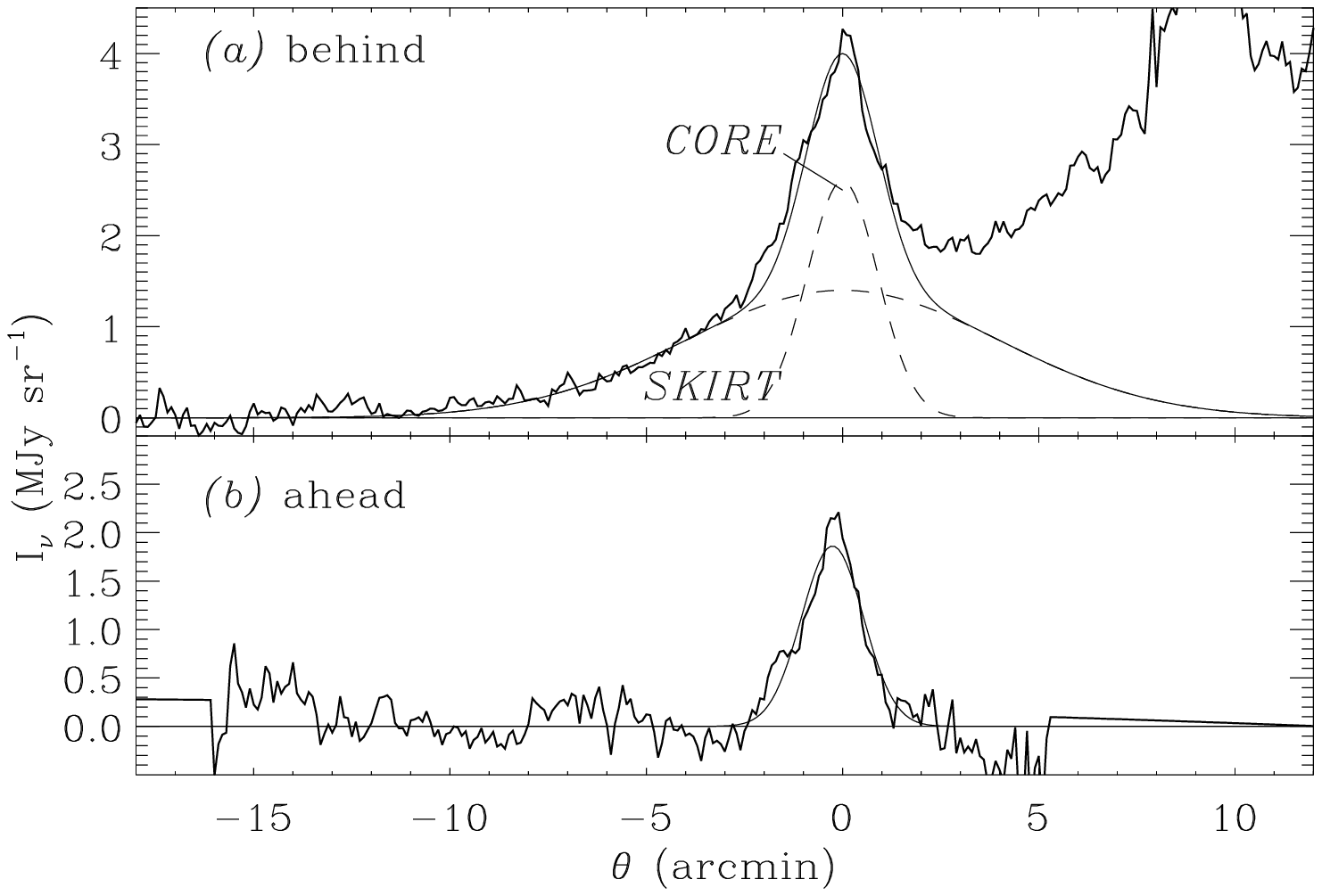}} 
\figcaption{\it
Profile of 2P/Encke dust trail on July 14, 1997 made from {\it ISO}
observations. The upper panel {\it (a)} shows a profile using only
the portion of the trail observed behind the comet, while panel {\it (b)}
shows a profile using only the portion of the trail ahead of the comet.
Gaussian fits to the profiles are shown by solid curve.
Both profiles are cleanly resolved, and the profile behind the comet
has a significantly irregular shape, which was fitted with two Gaussians.
each of which is shown as a dashed curve. The full widths at half maximum
intensity are $26^{\prime\prime}$ and $54^{\prime\prime}$ for the two
Gaussians.
\label{trailprof_encke}}
 
\begin{table}[tbh]
\caption{\it Brightness profiles of Encke dust trail}\label{proftab} 
\begin{center}
\begin{tabular}{lcccccrc} \hline\hline
\multicolumn{1}{l}{Year} & \multicolumn{2}{l}{Location$^a$} & 
\multicolumn{1}{c}{Wavelength} &
\multicolumn{1}{c}{Peak Brightness} & \multicolumn{2}{c}{Width (FWHM)} & 
\multicolumn{1}{c}{Optical depth} \\ \cline{2-3}\cline{5-6}
\multicolumn{1}{l}{ } & \multicolumn{1}{l}{$\Delta MA$ ($^\circ$)} & 
\multicolumn{1}{c}{($10^3$ km)} & 
\multicolumn{1}{c}{$\lambda$($\mu$m)} & 
\multicolumn{1}{c}{$I_\nu$(MJy~sr$^{-1}$)} & 
  \multicolumn{1}{c}{$W$($^\prime$)} & \multicolumn{1}{c}{($10^3$ km)} &
  \multicolumn{1}{c}{$\tau$} \\ \cline{1-8}
1997 & $+0.014$ & 150  & 12 & 2.6 & 2.1 & 23  & $9.5\times 10^{-9}$ \\
     &          &      & 12 & 1.4 & 9.4 & 102 & $5.1\times 10^{-9}$ \\
1997 & $-0.014$ & -150 & 12 & 1.9 & 1.9 & 20 & $6.8\times 10^{-9}$ \\ 
\\
1983$^b$ & $+52.8$ & & 60 & 0.06 & 4.7  & 680 & $1.3\times 10^{-9}$ \\ 
\hline
\end{tabular}
\end{center}
{\small\it \noindent $^a$ Location of the trail profile relative to the nucleus.
$\Delta MA$ is the difference in mean anomaly of the trail particle with respect to 
that of the nucleus. A positive value means a particle is behind the comet,
and a negative value means a particle is leading the comet.
The next column gives the distance to the nucleus in $10^3$ km.
\noindent $^b$ from Sykes and Walker (1992); averaged over a number of scans
whose average delta mean anomaly was $52.8^\circ$}
\end{table}

Much of the ISOCAM image is filled with emission from Encke dust. 
To check whether the background emission in our
image is partially due to a very extended distribution of dust spilling
over the edges of the image, we use the {\it COBE}/DIRBE zodiacal light
model (Kelsall et al. 1998) and the ISOCAM filter transmission 
(P\'erault et al. 1994) to estimate the zodiacal
light in our filter and on our observing date to be $13\pm 2$ MJy~sr$^{-1}$.
Because our observation was at a relatively low galactic latitude,
we also expect some emission from the diffuse interstellar medium.
Using the DIRBE-corrected {\it IRAS} dust template of Schlegel et al.
(1998) to get the interstellar surface brightness at 100 $\mu$m wavelength,
and a nominal interstellar dust spectrum (Reach and Boulanger 1998),
we estimate the brightness of interstellar dust in our image to be
$1.2$ MJy~sr$^{-1}$. Thus the total predicted background for our
image is $14\pm 2$ MJy~sr$^{-1}$.
The faint regions in our image, near the bottom, have a brightness of 
$13.7\pm 2$ MJy~sr$^{-1}$, which is entirely consistent with the
predicted background. Therefore, it appears that our image does
extend far enough to catch the southern edge of the
emission from Encke dust. The dust trail clearly extends off the
image toward the west, and the coma clearly extends off the image
toward the northwest.

\section{Nucleus and inner coma}

Figure~\ref{encke_nuc} shows a
close-up of the inner coma made using data from a single raster leg,
with logarithmic contours. The individual pixel that pointed at the
nucleus was saturated, meaning that the brightness of the $1100
\times 1100$~km area centered on the nucleus cannot be determined. An
approximate fit to the coma surface brightness profile between
$6^{\prime\prime}$ (1100 km) and $6^\prime$ (65,000 km) is
\begin{equation} 
I_\nu \simeq 70 (\theta/30^{\prime\prime})^{-1}
\,\,{\rm MJy~sr}^{-1}, 
\label{eg:coma} \end{equation} 
where $\theta$
is angular distance from the nucleus. Note that the coma is highly
asymmetric at $\theta>30^{\prime\prime}$, 
and this equation only gives the azimuthally-averaged
falloff of brightness with distance from the nucleus. 
(Also, the coma profile flattens close to the nucleus; a better fit
to the coma from $1^\prime$ (11,000 km) outward has a
radial power-law exponent of -1.15.)
A background
of 13.7 MJy~sr$^{-1}$ (the brightness in the edges of
the image)
due to zodiacal light and diffuse interstellar medium, was
subtracted from the image (see previous section).
Using the dust temperature of 270 K, measured by
Lisse {\it et al.} (2000) with broad-band 5--100 $\mu$m ISOPHOT
observations taken only 4 days later than our observations, the
projected optical depth of the coma is 
\begin{equation} \tau \simeq
2.6\times 10^{-7} (\theta/30^{\prime\prime})^{-1}. 
\label{eq:comatau}
\end{equation}
Because the profile is flatter than $\theta^{-2}$, the integrated
flux from the comet is not strongly dominated by the inner coma. The
dust coma of Encke was very bright, with an integrated flux within
$6^\prime$ (65,000 km) diameter of 40 Jy.

\centerline{\psfig{figure=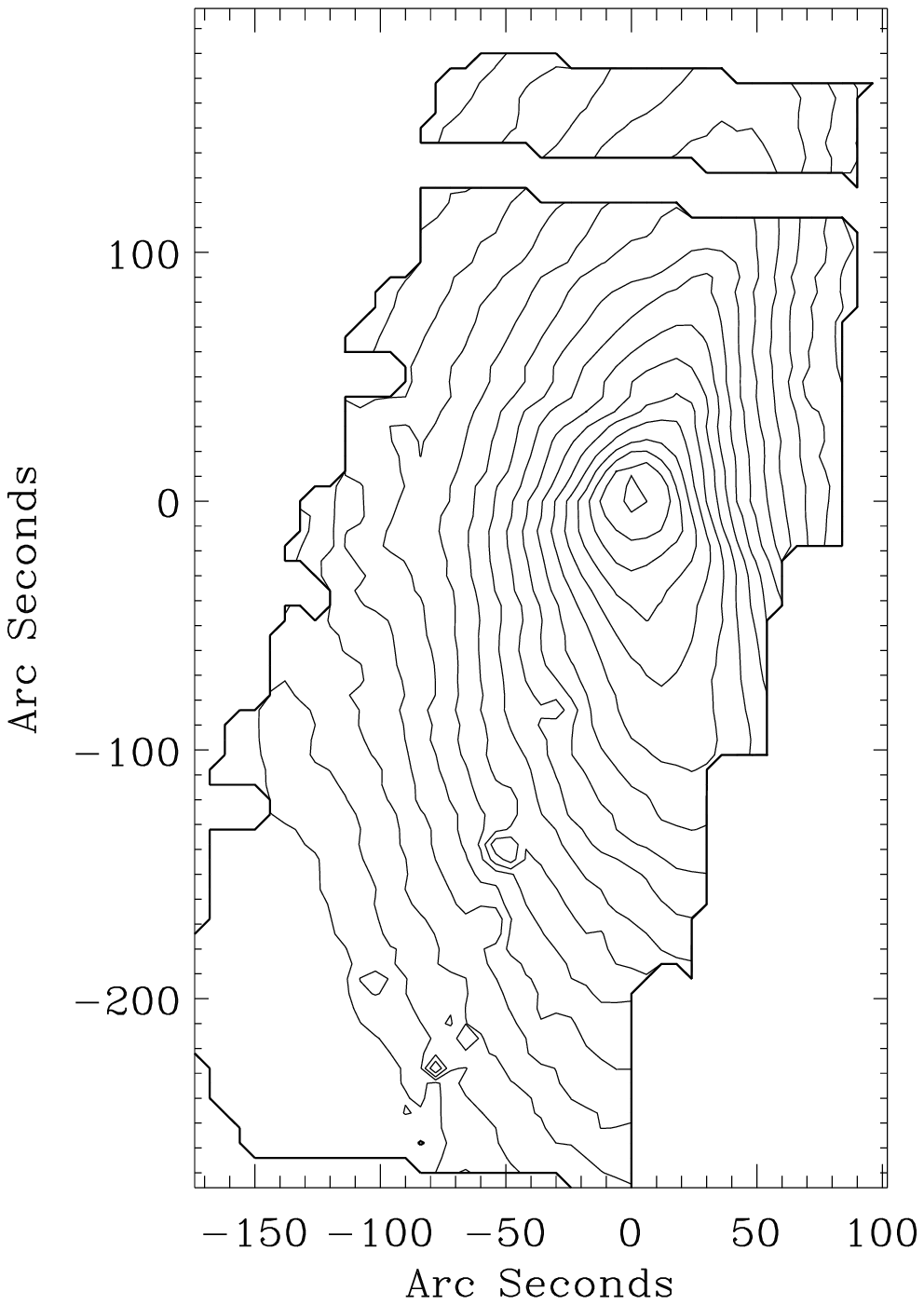}} 

\figcaption{\it
Close-up view of the coma of 2P/Encke from
the 1997 {\it ISO} observation. Only data from 3 pointings within a
single raster leg were used, to minimize smearing due to comet
motion; the horizontal band of no data near the top is due to the
dead column in the ISOCAM array. The contours range from 4 to 126
MJy~sr$^{-1}$, with a factor of 1.259 between each contour. The axes
are labeled with distance from the nucleus in arc seconds.
\label{encke_nuc}}

The radius of the nucleus of P/Encke was recently measured by
Fernandez et al. (2000) using ESO and HST observations; they found
$R=2.4\pm 0.3$ km. Using our observations
combined with the ISOPHOT observations, we can make an independent
estimate of the size of the nucleus. 
The single ISOCAM pixel containing the nucleus was
saturated, which means that the flux in that pixel exceeded 2 Jy. If
we extrapolate the coma surface brightness inward using
eq.~\ref{eg:coma}, the coma contributes about 0.9 Jy to the central
pixel. The remaining $> 1.1$ Jy/pixel could be due to the nucleus or
to enhanced inner coma brightness (relative to our extrapolation).
For a point source, about half of the flux falls in the central pixel
(based on integrating the ISOCAM point spread function; Okumura
1998), so the lower limit to Encke's flux is 2 Jy. An upper bound on
the nuclear flux is set by an unsaturated measurement of the flux in
a $3^\prime$ aperture using ISOPHOT (Lisse {\it et al.} 2000). 
Interpolating
between the ISOPHOT 10 $\mu$m and 12.8 $\mu$m fluxes, the flux at our
wavelength would be 19 Jy. Integrating our ISOCAM image over a
$3^\prime$ diameter aperture centered on the nucleus (but excluding
the central, saturated pixel), the flux is 21 Jy.
From controlled observations of
calibration stars, the relative calibration between ISOCAM and ISOPHOT
is accurate to $\sim 10$\% (Moneti 1998); also, the observations themselves
have $\sim 10$\% uncertainties. Therefore the nuclear
flux from the ISOPHOT-ISOCAM difference is $-2\pm 3$ Jy, which we can
interpret as an upper limit of 4 Jy. The approximate confidence
interval of nuclear fluxes, 2--4 Jy, constrains the radius of P/Encke
as follows. If the nucleus were a absorbing, rapidly-rotating sphere
with a subsolar latitude of zero, its equatorial temperature would be
260 K, and the radius must be in the range 1.5--3 km. If the nucleus
were a slow-rotating sphere with a single temperature on the sunlit
hemisphere and zero on the other, the sunlit hemisphere
temperature would be 307 K, and the radius would be in the range 1--2 km. 
Thus the radius of P/Encke
must be in the range 1--3 km. This compares well with the
direct radar measurement of Encke's radius, which yielded a
confidence interval 0.5--3.8 km (Kamoun {\it et al.} 1982).
Our limit is similar to previous limits determined from infrared and
optical observations (Campins 1988; Gehrz {\it et al.} 1989; Luu and Jewitt 1990;
Fernandez 2000).

\def\extra{The detection of a very bright and extended coma demonstrates the
dramatic increase in sensitivity possible with space measurements at
infrared wavelengths. Ground-based observations with the ESO 3.6-m
telescope only 1 week later than our observations revealed only a
point-like detection of comet Encke (Lisse {\it et al.}, in preparation). 
This ground-based detection has comparable contributions from
the inner coma and the nucleus,
and it was unable to reveal the extended
dust emission. The great advantage of space-based observations for
diffuse emission is that the bright atmospheric background is not
present, obviating the need to chop rapidly between sky and reference
position. The bright and extended coma of comet Encke was erased by
the sky-chopping observing technique and swamped by the bright
thermal emission of the Earth's atmosphere.}

\section{Dynamics of the Encke Dust Trail}

\subsection{Dynamical principles}

Space-based infrared observations of short-period comets reveal a more
extensive dust environment than what is easily observable from 
ground-based telescopes. To understand the origin of this extended
dust environment, we need to compare it to predictions for particles
of a given ejection velocity ($v_{ej}$) and size (parameterized
by $\beta$). A dynamical analysis of
cometary particles yields {\it only} $\beta$ and $v_{ej}$, from which may
be inferred the size (and mass).
Assuming that the particles are spherical with radius $a$ ($\mu$m) and
mass density of $\rho$ (g~cm$^{-3}$), 
and they are large compared to the wavelength of sunlight ({\it i.e.} $a\gg 1$),
then 
\begin{equation}
\beta = \frac{K}{\rho a},
\label{eq:beta}
\end{equation}
where $K=5.7\times 10^{-5}$ g~cm$^{-2}$ (Burns et al. 1979). 
Note that the particles size for a given $\beta$ is rather uncertain,
because it depends on the particles' mass density and geometry.
Collected interplanetary dust particles and chondritic meteorites appear
rather solid, with $\rho\simeq 1-3$ (Corrigan et al. 1997), but
cometary dust may be rather porous and irregular (cf. Lien 1992).

\centerline{\psfig{figure=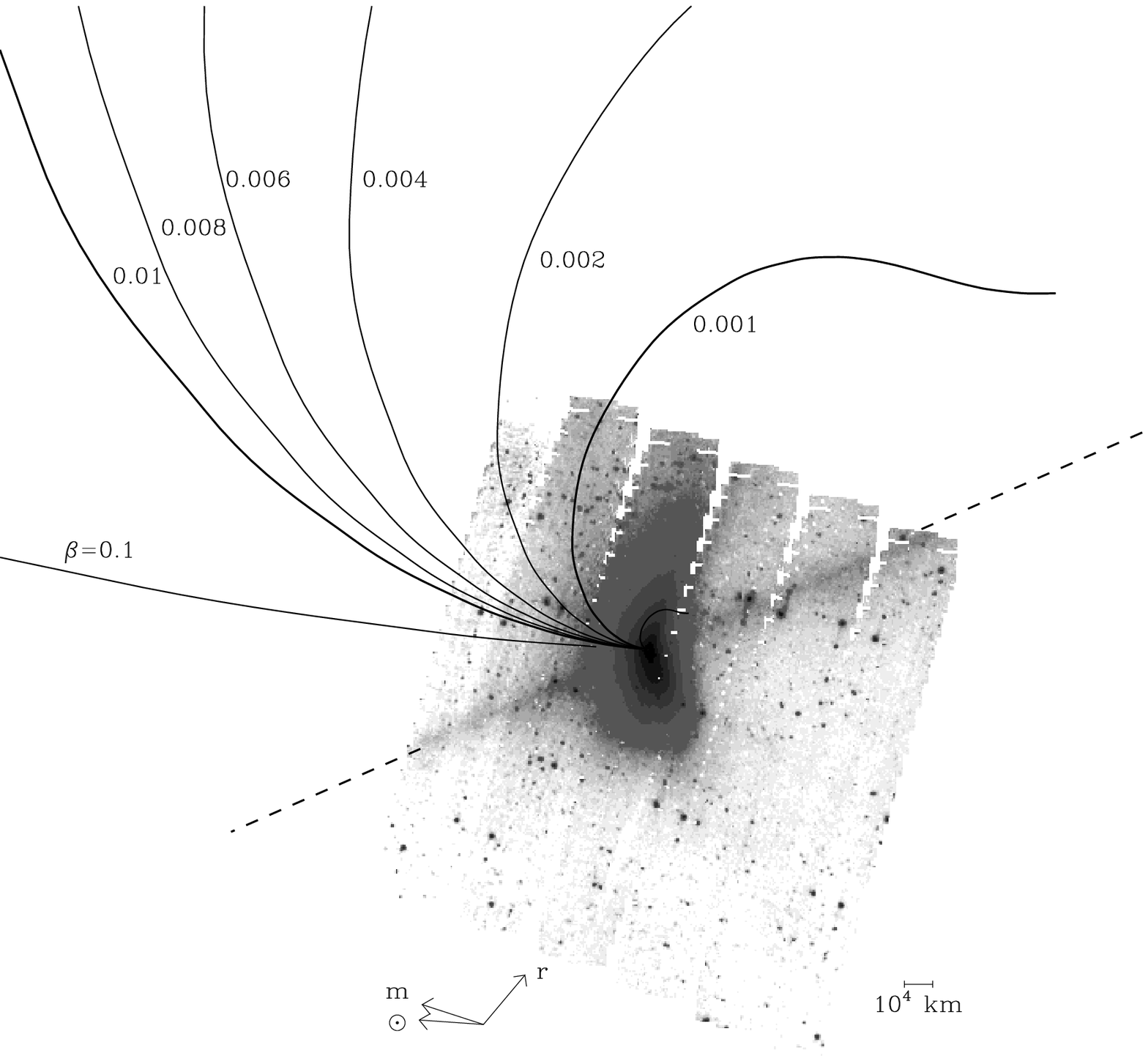}} 
 
\figcaption{\it Trajectories (syndynes) of particles with several different
values of $\beta$ (as labeled), 
ejected from the comet over the 260 days prior
to our ISOCAM observation. 
The short,unlabeled syndyne that stays close to the nucleus is for
a particle with $\beta=10^{-4}$.
The syndynes are overlaid on a greyscale
version of Figure~\ref{isoshift}.
The dotted line shows the projected direction of Encke's orbit.
The position angle of the sun is $86^\circ$ east of north.
The outlined field of view covered by the ISOCAM image in the comet's 
frame $30^\prime \times 33^\prime$ on the sky, corresponding to
330,000 $\times$ 360,000 km.
\label{syndynes}}

If the ejection velocity is assumed to be zero, then lines of
constant $\beta$ (syndynes) can be used to determine the range of
$\beta$s ejected by a comet. Figure~\ref{syndynes} shows the syndynes overlaid on
a greyscale version of the ISOCAM image. 
This figure can be used to define the
canonical three main parts of the comet: coma, tail, and trail.
Once the dust is ejected it will end up in either
the tail or the trail, and the `dust coma' is only a transition region.
We consider the dust coma to be a roughly spherical
region around the nucleus whose radius is equal to the distance at
which sunward emitted dust with intermediate values of $\beta$ `turns around' in
the reference frame of the comet due to radiation pressure.
For our observations of comet Encke, the `dust coma' is the peanut-shaped
region of intense emission with a diameter of about 
$9^\prime$ ($10^5$ km) centered on the nucleus.
Dust with larger values of $\beta$
would form a comet tail; 
particles ejected at perihelion will have unbound orbits if 
$\beta>(1-e)/2$ (Burns et al. 1979),
which means $\beta>0.075$ for Encke.
Based on Figure~\ref{syndynes} and the
more detailed dynamical models presented below, there is no evidence 
for a dust tail due to small particles (less than 200 $\mu$m in radius)
in comet Encke. 
The syndynes for particles with very
small $\beta<10^{-4}$ are not shown in Fig.~\ref{syndynes}
because they crowd into the dust trail, as is found in the simulations below.

\def\extra{
A dust trail can be best understood in terms of the orbital elements
of the dust particles. Each grain ejected from the nucleus is 
in its own orbit around the sun. For grains with high $\beta$, the orbit is
hyperbolic, and the grain escapes from the solar system.
Particles ejected at perihelion will have unbound orbits if 
$\beta>(1-e)/2$ (Burns et al. 1979),
which means $\beta>0.075$ for Encke. These particles travel 
essentially radially outward on hyperbolic orbits and will not be part
of the curved Encke dust tail that we observed.
The orbits of all other grains can be
described using the normal six orbital elements for an elliptical
orbit.
Without radiation pressure, grains ejected isotropically from the
nucleus would eventually create a torus of dust centered on the orbit
of the comet;
the thickness of the torus would depend on the ejection
velocity of the grains. 
If the effects of radiation pressure on the
orbital elements of the grain are small compared with the changes due
to the ejection velocity, then the grains will remain within the
torus. If the effects of radiation pressure on the orbital elements
of the grains  are large compared to with the effects of the ejection
velocity, then the dust grains will move out of the torus  to form
the tail of the comet. 
Clearly, however, the tail actually `merges' with the trail for
some intermediate values of $\beta$, making the distinction between tail
and trail somewhat arbitrary.
Some of these intermediate-sized particles
become part of the zodiacal dust cloud, either directly or after
being temporarily trapped into resonances with the planets (Liou and Zook 1996).

It is clear that the distribution of dust around comet Encke is only
vaguely approximated by the syndynes (Fig.~\ref{syndynes}).
However, since P/Encke is known
to have strong asymmetries in its dust ejection (Sekanina 1988a,b)
this is to be expected. Figure~\ref{syndynes} illustrates that
the dust environment is dominated by particles with 
$\beta < 10^{-2}$.
In contrast to P/Encke, more active comets like P/Halley and
C/Hale-Bopp have dust tails dominated by particles with
$\beta >1$ 
(Lien {\it et al.}, in preparation).
Additionally, for more active comets,
the dust extends in the sunward direction and
ends in a smooth parabolic envelope. P/Encke
evidently does not behave in this fashion, because
the distribution of particle sizes is very different for
P/Encke as compared to P/Halley or C/Hale-Bopp.
}

\subsection{Numerical simulations}

To better understand the observed shapes of the coma and trail,
we created a series of simulations using a Monte Carlo dynamical model 
(cf. Campins et al. 1990). 
Our simulations assume a differential size distribution
\begin{equation}
 \frac{dn}{d\beta} \propto \beta^{-k},
\label{sizedist}
\end{equation}
ranging  from $10^{-2}$ to $10^{-6}$, with $k=1$ as the initial guess.
The ejection velocity depends on size and heliocentric distance as
\begin{equation}
v_{ej}= 1.35 \beta^{0.5} R^{-0.5}\,\,{\rm km~s}^{-1}
\label{velbeta}
\end{equation}
(cf. Whipple 1951).
Lisse {\it et al.} (1998) found that this velocity distribution
reproduced the COBE observations of
C/Austin, C/Levy, and C/Okazaki-Levy-Rudenko. Over the range of 
$\beta$s used in the simulations, our ejection velocities
(eq.~\ref{velbeta}) reproduce Sekanina's ejection velocities (1988a,b)
to within 10\%. 

\centerline{See 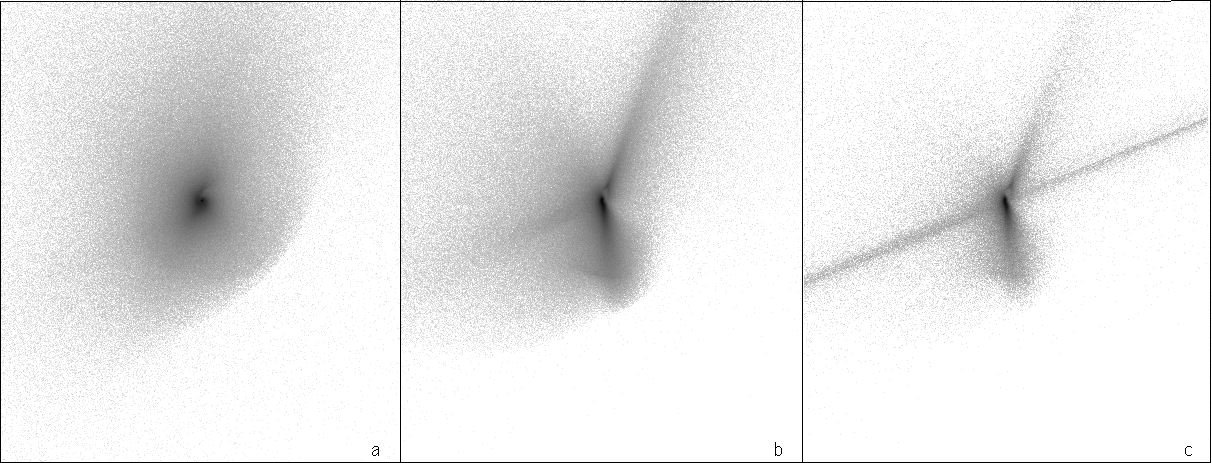}
\figcaption{\it
Monte Carlo simulations of dust emitted since the 1995 aphelion for 
{\it (a)} isotropic emission from the sunward hemisphere; and
{\it (b)} jet emission from a rotating nucleus 
(see text for jet latitude and pole orientation).
Panel {\it (c)} is the same simulation as panel {\it (b)}, except the
dust emission begins at the previous (1992) aphelion. 
For each panel in this Figure and Figs.~\ref{figfour}--\ref{figsix},
the intensity in each panel is proportional to the logarithm 
of the flux from particles in each pixel, and each panel
shows a region approximately 440,000 km on a side, with celestial N
upward and E to the left. The orientation in this and subsequent figures
is the same as in Figs.~\ref{isoshift} and~\ref{syndynes}.
 \label{figthree}}

Figure~\ref{figthree} shows
the results of three simulations. 
Each panel shows the number of particles per pixel weighted by the
time a particle spends in that pixel. The images were smoothed to
$15^{\prime\prime}$ resolution to improve their appearance.
The first simulation, shown in panel Figure~\ref{figthree}a,
assumes isotropic emission over the sunlit hemisphere.
The dust production rate scales 
with heliocentric distance as $\dot{M_{d}} \propto R^{-2}$,
with dust ejection beginning at aphelion ($R=4.1$ AU; October 1995).
The isotropic emission model in Figure~\ref{figthree}a
fails to reproduce any of the coma structure 
observed in our ISOCAM image. This confirms the previous work 
that demonstrated that Encke's mass loss is anisotropic, based
on the non-gravitational forces and orientation of the
visible fan-like tail (Sekanina 1986, 1988a,b), that
showed that the 
visual appearance of Encke's fan-like dust tail
could be understood in terms of a vents on the northern and southern
hemispheres of a rotating and precessing nucleus.
The isotropic emission model with smooth $\dot{M}$ also fails to reproduce
the observed time-variation of the OH production rate (A'Hearn et al. 1985).
An active area on
the northern hemisphere was suggested by Fanale and
Salvail (1984) to explain observed variations in OH production, and
by Newburn and Spinrad (1985) to explain an anomalous decrease in
dust loading of the coma near perihelion. 

The second simulation, shown in Figure~\ref{figthree}b,
assumes emission from a single jet at a latitude of $-15^\circ$ with
the north pole pointing at at RA=102.8$^\circ$, Dec=32.6$^\circ$.
This polar orientation corresponds to an obliquity of nearly
$90^\circ$, so that Encke's pole is in its orbital plane.
The rate of particle ejection from the jet was proportional to the
cosine of the zenith angle of the Sun.
A rotational period of 24 hours is used in this analysis. This period 
differs from the 15 hours found by Luu and Jewitt (1990). However,
our  results are completely insensitive to rotational periods of less
than a few days, because any rotational structure is smoothed out by
variations in $\beta$ and the low ejection velocities.
The single-jet, rotating model in Figure~\ref{figthree}b includes dust emission 
occurring since the
most recent aphelion (October 1995). This model reproduces the 
gross features of the ISO
image: the southern lobe of material in the coma, the `spike' to the
NW, and the orientation and width of the trail (illustrated in
Fig.~\ref{spikefig}). However, the extent of
the trail is not reproduced. Figure~\ref{figthree}c shows the
same model as Figure~\ref{figthree}b, but with dust ejection beginning at the
aphelion of the {\it previous} apparition (June 1992). Clearly,
reproducing all of the observed structures requires not only that the
dust is ejected anisotropically, but also that the dust is emitted
over multiple orbits.

To better understand the types of particles that give rise to the
dust coma and trail, we repeated the rotating jet model used in 
Figure~\ref{figthree}b 
for particles in four separate decades of $\beta$: 
$10^{-2}$--$10^{-3}$; $10^{-3}$--$10^{-4}$; $10^{-4}$--$10^{-5}$; and 
$10^{-5}$--$10^{-6}$. Figure~\ref{figfour}
shows the simulated images for particles in these four size ranges.
The ISOCAM image must be
produced by particles with $\beta < 10^{-3}$, because the higher-$\beta$
particles create a diffuse coma that does not look at all like the one we 
observed.
The bright lobe of emission extending $\sim 2^\prime$ to the southwest
of the nucleus, and the `spike' to the northwest, are due to the projection of
the expanding material from the jet. The `gap' in the images
for intermediate-$\beta$ particles is created when the jet
is pointed away from the Sun, near perihelion, due to the tilt of
Encke's rotational pole---that is, when it becomes `winter' for the jet
(late April 1997). 
The `turn-around' seen coming from the `spike' is due to the first
material ejected just after the comet turned on again during this most recent
apparition (mid June 1997). 
Figure~\ref{enckeday} shows the locations of Encke and Earth for these
events.
All of the dust observed in these simulations
eventually ends up in the dust trail of
comet Encke. 

\centerline{See 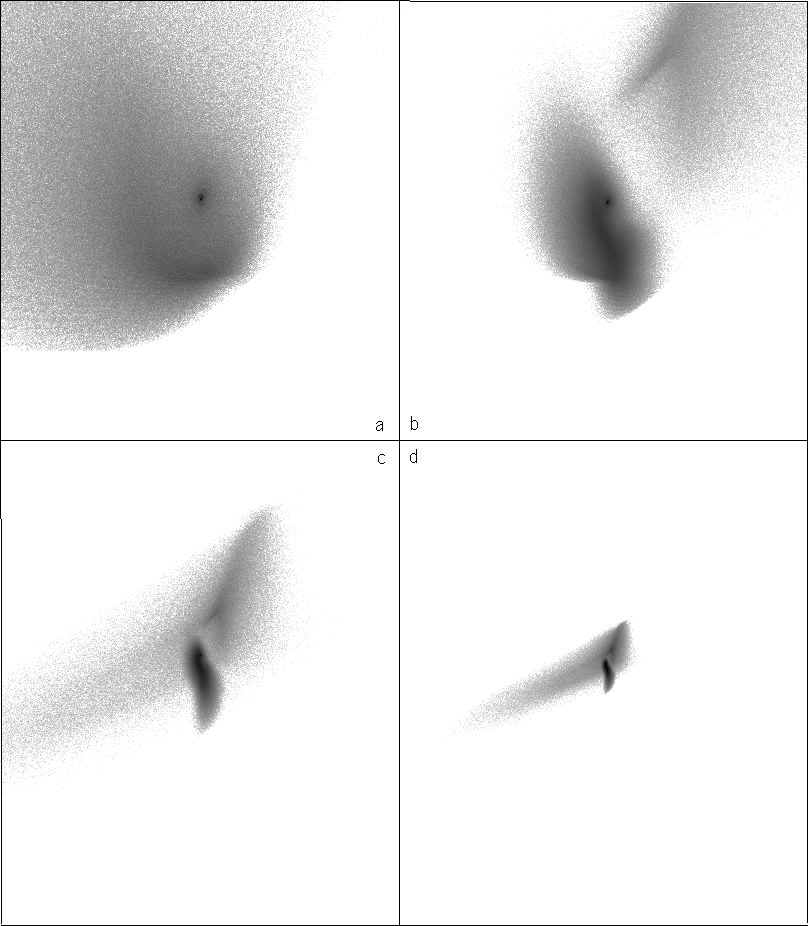}
\figcaption{\it
 Monte Carlo simulations of emission of particles emitted since the
 1995 aphelion (same model as Fig.~\ref{figthree}b)
split into four decadal ranges of $\beta$: 
{\it (a)} $10^{-2} \ge \beta \ge 10^{-3}$; 
{\it (b)} $10^{-3} \ge \beta \ge 10^{-4}$; 
{\it (c)} $10^{-4} \ge \beta \ge 10^{-5}$; and 
{\it (d)} $10^{-5} \ge \beta \ge 10^{-6}$.  
Panel {\it (a)} is so different from the observed coma shape that we
conclude that particles in that size range do not contribute to the observed
emission. 
The `gap' in the NW `spike' occurs because the jet turns off for a period of 
time near perihelion (see text for additional details).
The dust trail has not yet formed in these simulations.
 \label{figfour} }

Only the oldest and largest particles from our simulation of the present 
apparition of Encke are close to joining the dust trail. 
Thus, we conclude that the core of the dust trail contains mostly
particles emitted during {\it previous} apparitions. To
reproduce the entire dust trail as seen in our image, we ran the
same model as Figure~\ref{figthree}c for the same 4 particle size
ranges as Figure~\ref{figfour}.
Dust production begins in
June 1992, at the aphelion of the {\it previous} apparition.
Figure~\ref{figfive} shows a number of important features. 
First, even including particles from the previous apparition, particles
with $\beta>10^{-3}$ do not contribute significantly to the ISOCAM image.
Second, Figure~\ref{figfive} shows that the particles that extend off the
northwest edge of the ISOCAM image most likely have intermediate size:
$10^{-4} > \beta > 10^{-5}$. 
Third, the core of the dust trail is produced by very large particles. 
The width (FWHM) of the simulated trail in Fig.~\ref{figfive}c is 
$4^\prime$, while the width of the trail in Fig.~\ref{figfive}d is 
$2^\prime$. Fig.~\ref{figfive}d is a very good match to the `core'
of the dust trail in Figure~\ref{trailprof_encke} and Table~\ref{proftab};
therefore, the core of the dust trail is produced by particles with
$\beta< 10^{-5}$. The `skirt' of the dust trail, which appears only 
trailing the nucleus, is produced by particles with a range of sizes,
with a typical value $\beta\sim 10^{-4}$.

\centerline{See 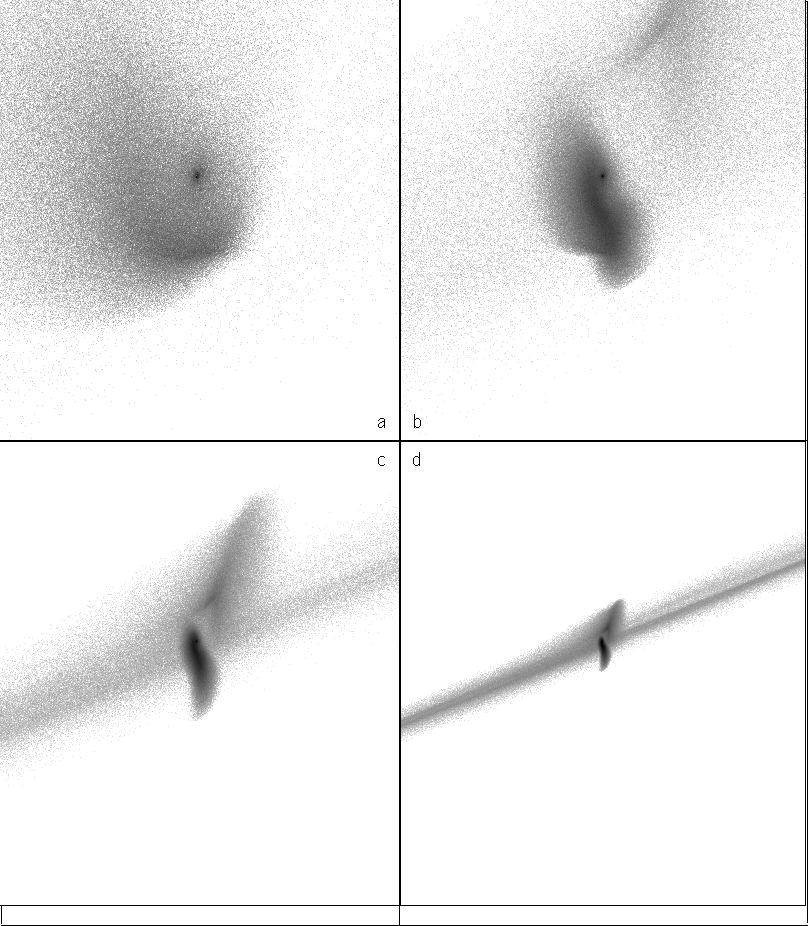}
\figcaption{\it
Same as Figure~\ref{figfour}, but for emission starting at the
previous (1992) aphelion (same model as Fig.~\ref{figthree}c).
Dust is not fully entrained into the trail until after one complete orbit.
Panel {\it (d)} is a good match to the core of the dust trail,
allowing us to estimate the size of trail particles.
\label{figfive}}

The ISOCAM image shows, for the first time, cometary trail material {\it before} it 
actually enters into the trail. 
In principle, this allows us to determine the trail mass injection rate
unequivocally once we have the correct rotational and jet model.
Unfortunately, the model presented in Figure~\ref{figthree} is not
unique. Figure~\ref{figsix} shows the simulation of 
Figure~\ref{figthree}c along with a simulation
based on Sekanina's rotational model (two jets with latitudes of
$-75^\circ$ and $+55^\circ$ and a pole position of RA$=206^\circ$,
Dec$=+3^\circ$ which corresponds to an obliquity of $70^\circ$). 
In the Sekanina model, the  NW `spike' is created solely by the $-75^\circ$ jet,
which turns off just before perihelion, and the southern lobe is
created solely by the $+55^\circ$ jet. In our model both
features are created by the same jet.
Sekanina's analysis assumed that the bisection
of the observed visible emission fan indicates the instantaneous
projection of the spin vector. 
This assumption is only correct if the dust is relatively `young.' 
If the fan
is created by particles with $\beta < 10^{-3}$, as suggested by our
results, then the youngest
dust particle would be over 1 month old at the outer edge of the
wedge. Since most of the observations were made around perihelion,
there can exist a large change in the projection of the spin axis
over the course of 1 month, in which case the instantaneous
projection of the spin axis will {\it not} coincide with the
bisection of the observed dust wedge.
Therefore it is very difficult to make an accurate measurement
of the orientation of the rotational pole.
Neither our model nor Sekanina's model are perfectly matched to 
the observations. 
However, given the relatively large phase
space of possible solutions, it is interesting that such simple
models can match the ISOCAM observations as well as they do. 
Taken together, our work
and that of Sekanina show that a high
obliquity is required to interpret
both infrared and optical observations of dust from Encke.

\centerline{See 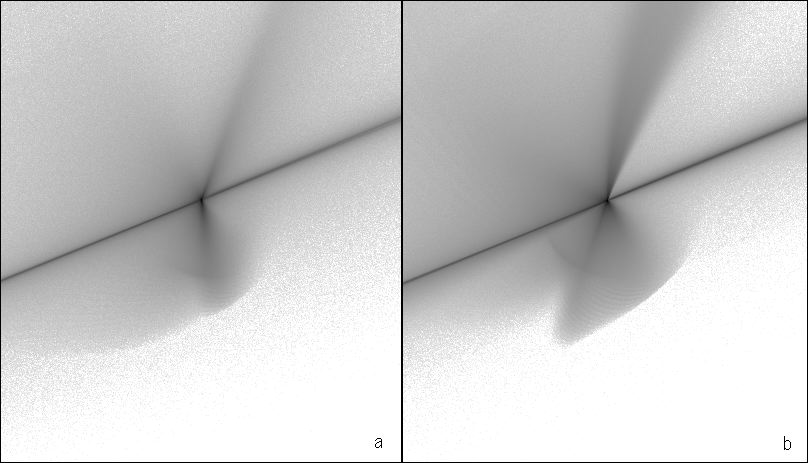}
\figcaption{ \it
Monte Carlo simulations of dust emission comparing {\it (a)} 
our single-jet rotational model with {\it (b)}
the two-jet rotational model of Sekanina
(1988a, 1988b) for emission beginning at the previous (1992) aphelion.
These models are morphological
only, and do not incorporate either a $\beta$ distribution or the
emissivity as a function of $\beta$. The similarity of the 
predictions of the two models shows that the visible and infrared
observations can both be fitted with a similar model of emission
from a spinning nucleus tilted into the orbital plane.
\label{figsix}}

\subsection{Model results}

Using the dynamical simulations, we constrained the particle sizes
through their force ratio $\beta$ (Eq.~\ref{eq:beta}).
Each particle was assigned a surface brightness equal to a Planck
function at $T=270$~K (see above) multiplied by the solid angle of the particle
as seen from Earth. Then the production rate and $\beta$ distribution were
adjusted in order to reproduce the observed surface brightness.
The production rate was adjusted by making simulations for particles
ejected over short intervals and determining how their contributions
add to make the total image. 
The initial guess for the production rate was determined by choosing a 
narrow range of $\beta$ and calculating the surface brightness of dust
ejected over a small range of heliocentric distance. Dividing the
observed image by the initial-guess simulation yields an estimate of the 
heliocentric dust production rate. 
The $\beta$ distribution 
was also adjusted by making simulations for size ranges and determining
how their relative contributions add to make the total image.
The initial guess for the $\beta$ distribution
was obtained by choosing a heliocentric distance where there is a good
separation between the effects of $\beta$ and production rate, then
calculating a simulated image for a small range of $\beta$ at that heliocentric
distance. 
These adjustments were made step by step, and our final solution is not unique;
for example, the orientation of the pole and the jet are correlated
with the production rate. However, the salient features of the model are robust.
The simulations were normalized by the total number of particles emitted over
the simulation multiplied by the production rate and $\beta$ distribution
functions. 
Because the model
assumes a single jet which turns on and off with season, the resultant
production rate is a complex function of heliocentric distance.

We found the $\beta$-distribution index from 
equation~\ref{sizedist} to have a value $1.3<k<1.6$ for the range
$10^{-7}<\beta<10^{-3}$, with no particles with larger $\beta$.
This is not a unique solution for the size distribution, and a
much more exhaustive modeling effort would be able to
determine the size distribution in more detail---including the 
higher-$\beta$ particles.
For our size distribution, both the mass and the surface area are in
relatively large particles. The mass is in the very largest particles
and the area is in particles with $\beta\sim 10^{-3}$.
The ISOCAM observations and our model clearly
shows a paucity of the smallest particles (with high $\beta$). This
is not due to physical optics alone, which predicts that
particles with 
$2\pi a/\lambda<1$ will not emit efficiently.
For our observation $\lambda=11.5$~$\mu$m, so only particles smaller than 
2 $\mu$m radius, or $\beta\simeq 0.3$, will emit inefficiently due
to physical optics.
But we find a paucity of particles with $\beta>10^{-3}$, which
corresponds to $a<600$ $\mu$m. If physical optics were the reason for
the observed lack of high-$\beta$ particles, then the relation between
size and $\beta$ would have to be modified drastically. 
Alternatively, there could be a real lower limit to the size of particles
ejected from Encke. In either case, our inferred $\beta$ distribution
differs significantly from those found for other comets such as
P/Halley, for which the mass distribution is dominated by large particles
while the surface area (and the infrared emission) is dominated by
much smaller particles (McDonnell et al. 1991). We find that
the infrared emission from P/Encke is dominated relatively large
particles, so the size distribution
for Encke must be more shallow than that for Halley.

Using the numerical simulations that best match the ISOCAM image,
we find that the total mass emitted since aphelion is within the range
2--6$\times 10^{13}$ g (assuming particles with mass density $\rho=1$ g~cm$^{-3}$).
The range in values comes from the different $\beta$ distributions that 
produced acceptable matches to the observed surface brightness of the comet.
The model we present here is not unique, and it could be improved by
optimizing the parameter values to match the observations in detail.
Therefore, our derived mass loss rate remains uncertain, although
the factor of 3 range is our best estimate.

To estimate the contribution to the trail from the previous apparition, we ran
the numerical model for approximately 1.5 orbits, starting from the aphelion
before the previous apparition. In the core of the trail, the model brightness
is only about 10\% of the observed brightness. This suggests the
trail is at least 10 orbits old, which is consistent with the trail
age estimated from its total extent (Sykes and Walker 1992). The mass of the
trail is therefore at least 2--6 $\times 10^{14}$ g, comparable to the
estimate of $1.6\times 10^{14}$ g estimated from the {\it IRAS} 
observations (Sykes and Walker 1992).

As a check, a model-independent estimate for the mass of the 
portion of the dust trail present in the {\it ISO} image
can be measured from the observables as follows. If we assume the trail
is a cylinder oriented perpendicular to the line of sight,
with cross-sectional angular diameter $W$ and length $L$,
the distance from the Earth is $\Delta$, particle density is $\rho$, and
the optical depth through the trail is $\tau$, then
\begin{equation}
M_{trail} = \frac{\pi}{3} \Delta W L \rho a \tau,
\label{mtrail}
\end{equation}
where $a$ is the particle radius in the dust trail. 
This equation assumes that both the mass and emitting surface area are
produced by large particles of effective size $a$.
Because only the combination $\rho a$ determines the trail mass,
and the dynamics constrain $\beta\propto (\rho a)^{-1}$, the masses that
we derive here are independent of the mass density $\rho$ of the particles,
and the masses are independent of whether the particles are fluffy or compact.
We can set a generous upper limit $\beta < 10^{-4}$ by comparing the dynamical
simulations to the morphology of the dust trail and coma.  
Using the brightness profile fit from Table~\ref{proftab} we find a trail mass 
within our image
$M_{trail} \gg 1\times 10^{12}$~g, with
comparable amounts coming from the core and skirt. 
This trail mass is a drastic underestimate for at least three reasons.
First, the lower limit to $\beta$ determined from the dynamical simulations is
far form the typical values that create the observed dust trail; the particles
form previous apparitions are in the core of the trail and can have $\beta$s
that are more than an order of magnitude lower. 
Second, even if $\beta=10^{-4}$ particles were to dominate the brightness, larger
particles would also exist and could dominate the total dust mass if the 
size distribution is at least as shallow as was observed for P/Halley
(McDonnell et al. 1991).
Third, it is clear that the
trail extends off the edge of the ISOCAM image. Using the {\it IRAS} 
observations, Sykes and Walker (1992) found the Encke dust trail extends fully
$90^\circ$ of mean anomaly relative to the nucleus.
At the time of the {\it IRAS} observation, this corresponded to a length of 2 AU,
and at the time of our observation, the same mean anomaly would correspond to a length
of 5.5 AU along the orbit. Our ISO observation covered only 0.002 AU of the comet's orbit.
The actual mass of the Encke dust trail is at least an order of magnitude larger that 
our limit just due to this affect. 
These arguments suggest the trail mass is at least
$10^{14}$ g, which is consistent with the results of our numerical
simulations and provides a sanity check.

\section{Impact hazard in Encke's coma}

\def\old{
The impact hazard for a spacecraft passing through the extended
dust coma around comet Encke is not negligible. This is an
especially timely topic as the CONTOUR spacecraft is planning
to fly to within 200 km of Encke in 2003 (REF???). 
The amount of dust around the comet can be characterized by
the fluence (g cm$^{-2}$)
\begin{equation}
F = \int n m {\rm d}L,
\label{eq:fluence}
\end{equation}
where $n$ is the number of grains per unit volume, $m$ is the mass
of an individual grain, and $L$ is the path length;
the integral is along the trajectory of the comet.
With the infrared image, we can directly measure the optical
depth of particles, integrated along straight lines from the 
Earth through the comet's dust cloud:
\begin{equation}
\tau = \int n \sigma {\rm d}L,
\label{eq:tau}
\end{equation}
where $\sigma$ is the cross-sectional area of an individual grain.
In both of these equations, the `individual grain' refers to the
grains that produce the infrared emission. We cannot directly
constrain $m$ or $\sigma$ for the grains, but we do have an upper
limit on $\beta$ from the dynamics.
Combining equations~\ref{eq:beta}, \ref{eq:fluence}, and~\ref{eq:tau},
we find that the fluence 
\begin{equation}
F = \frac{4 K \tau}{3\beta},
\label{eq:flutau}
\end{equation}
assuming that $m/\sigma$ is the same throughout the coma.
A straight-line through the coma with impact parameter 200~km, 
the impact parameter goal for CONTOUR, 
would have an optical depth of $7\times 10^{-6}$,
based on equation~\ref{eq:comatau}.
Using this optical depth and the upper limit $\beta<10^{-3}$, we find
that the fluence $F>7\times 10^{-6}$ g~cm$^{-2}$. This lower limit
to the fluence is relatively insensitive to assumptions about the
particle properties; in particular, it is independent of the mass
density (hence, fluffiness) of a grain.

In the dust trail, there are relatively fewer particles than in the
coma, but they are larger. 
To estimate the hazard for a spacecraft passing through
the dust trail, we predict the number of impacts on a spacecraft
with area $A$:
\begin{equation}
N = \frac{A\tau}{\sigma} = \frac{A\tau \rho^2 \beta^2}{\pi K^2}.
\end{equation}
The predicted number of impacts is quite uncertain because it depends
too strongly on quantities that we cannot directly constrain.
For compact grains with $\rho=2.5$ g~cm$^{-3}$,
an a spacecraft with area $A=10$ m$^{2}$,
the expected number of impacts along a trajectory 
perpendicular to the trail is
$N \sim (\beta/10^{-3})^2$. The number of encounters for a spacecraft
approaching {\it along} the dust trail is of course significantly higher.
If we simply use the upper limit to $\beta$, we find only of order 1
or fewer impacts with trail particles are expected for a perpendicular
passage through the trail. In fact, the particles sizes in the core
of the dust trail are most likely significantly larger, with 
$\beta\sim 10^{-5}$. The impact probability with these large particles is
therefore rather small ($\sim 10^{-4}$);
however, each of these particles would be a very severe hazard,
with sizes comparable to baseballs. From this simple estimate, it
appears likely that large dust particles are a significant hazard 
to spacecraft approaching comet Encke. A more detailed analysis,
taking into account the different locations of different-sized particles,
is needed to determine whether these hazards represent an acceptable risk.
}

In November 2003, the CONTOUR spacecraft is scheduled to fly within 
200 km of Encke (Veverka 1999).
At this time the comet will be at heliocentric and 
geocentric distances similar to those when our ISOCAM observations 
were made.\footnote{Comet distances for the 2003 CONTOUR encounter were
calculated using the JPL Horizons program, 
{\tt http://ssd.jpl.nasa.gov/horizons.html}.}
Using a color temperature of 270 K, as measured 
measured by ISOPHOT (Lisse et al. 2000),
the optical depth at
an impact parameter 200~km is $\tau\simeq 7\times 10^{-6}$.
For spherical particles of a single size, the mass of particles 
encountered by a spacecraft (with area $A$) is
\begin{equation}
M_1 = \frac{4 K A \tau}{3\beta}.
\end{equation}
The encountered mass
is relatively insensitive to assumptions about the
particle properties; in particular, it is independent of the mass
density (hence, fluffiness) of the particles.
The number of particles encountered by the spacecraft is
\begin{equation}
N = \frac{A\tau \rho^2 \beta^2}{\pi K^2},
\end{equation}
which is sensitive to particle density as well as $\beta$.
Using the upper limit $\beta<10^{-3}$ for coma particles, we find
that a 10 m$^2$ spacecraft will encounter $> 0.05$ g of cometary particles
during its traverse. 
The expected number of particles encountered with
$\beta<10^{-3}$ is $N\sim 10^2$.

Using the size distribution from the numerical 
simulations (eq.~\ref{sizedist}), the encountered mass increases
relative to this lower limit to
\begin{equation}
M_{enc} = M_1 \frac{k-1}{2-k} f^{k-2} \frac{1-x^{k-1}}{1-x^{2-k}},
\end{equation}
where $x=\beta_{min}/10^{-3}$ and $\beta_{min}$ is the $\beta$ of the
largest particle (which dominates the mass). For example, for
$\beta_{min}=10^{-6}$ and $k=1.5$, the mass of cometary
particles encountered is $M_{enc}\simeq 1$ g. Larger particles
exist, but their spatial distribution is more confined. Thus, to
make an accurate assessment of the impact hazard, one would need
to compare the spacecraft trajetory and the particle distribution 
as a function of both particle size and position in the coma.
For comparison, the dust detectors on the Giotto spaceprobe passing through 
the coma of comet Halley in 1986 detected only 0.01 g, though an
estimated total mass of $\sim 0.15$ g penetrated the dust 
shields (McDonnell et al. 1986), at a distance of 2,200 km.
Also, several very large particles (1--50 mg each) must have impacted
the Giotto spacecraft to cause the observed attitude shifts 
(Curdt and Keller 1990).

In the dust trail core, there are relatively fewer particles than in the
coma, but they are larger. 
At high flyby velocities of tens of kilometers per second, such
large particles near the comet could represent a significant hazard to 
a spacecraft.
Using the trail optical depth
of $\tau\sim 1.4\times 10^{-8}$ and an upper-limit to $\beta<10^{-4}$,
there is $<0.1$\% chance of a spacecraft with an area of 10 m$^2$ 
hitting a trail particle, but each
particle is potentially catastrophic.

\section{Mass loss and meteoroid production of comet Encke}

The total mass of particles ejected by Encke in its 1997 apparition is 
2--6 $\times 10^{13}$ g. 
Dividing the mass of the nucleus (using $R=2.4$ km [Fernandez et al. 2000]
and a nuclear mass density $\rho_N$ in g~cm$^{-3}$) by the mass production
we infer for the 1997 apparition, the fraction of Encke's mass that was lost
in the 1997 apparition is
\begin{equation}
\frac{\Delta M}{M} = (0.1 - 1)\times 10^{-3} \rho_N^{-1}.
\end{equation}
For comparison, in his classic papers, Whipple (1950, 1951) found that 
the non-gravitational forces observed on comet Encke's orbit could be produced
by mass loss at a (recent) rate of $\Delta M/M=0.002$, with certain assumptions
about the nature of the mass loss. Our inferred mass loss rate is only smaller
by a factor of few, which is a remarkable confirmation
of the general principle of Whipple's comet model, in which the mass loss
rate depends on the velocity of the emitted particles and their angular
distribution (which determines the direction of the net momentum carried
away by the emitted material). A modern version of this model takes into 
account that the mass loss occurs primarily from small regions (`jets') 
whose illumination depends on season (Sekanina 1986, 1988a).
Further, a modern model should take into account that most of the
mass of comet Encke is lost in solids as opposed to gas, 
so there is a component of the non-gravitational force
due to the solids that could be somewhat different from that of the gas.
If the dust is lost through the same jets as the gas, as appears to
be the case from an analysis of the inner coma of P/Halley (Reitsema 1989),
then the only effect of our new dust-to-gas estimate is to increase
the amount of mass lost per unit ice sublimation. The non-gravitational
acceleration, which is the observable quantity, depends only on $\Delta M/M$.
Thus the main effect of increasing mass loss rate by a factor of 10--30 times
the gas mass loss rate is to increase the mass (and density) of the 
nucleus as inferred from non-gravitational forces.

The dust mass loss rate we infer from the dust trail observations
is indeed significantly larger than the gas mass loss rate. A'Hearn et al.
(1985) summarized observations the OH production along Encke's orbit,
and Sekanina (1986) provide a convenient fit. The total mass 
of sublimating H$_2$O is $\Delta M_{gas}=2.1\times 10^{12}$~g/orbit.
Thus the dust-to-gas ratio for comet Encke's mass loss 
is $\Delta M/\Delta M_{gas} \simeq 10$--30. Clearly our estimate is much
larger than the dust-to-gas mass ratios normally inferred for comets
based on optical observations. The fraction of the mass loss occurring
in non-volatile form is 91--97\%. For comparison, Whipple (1955) used 20\% as the
fraction of mass lost in non-volatile, meteoric, material. Based on the 
P/Halley encounter data from {\it Giotto}, the dust-to-gas mass ratio
was found to be $\sim 2$ (McDonnell et al. 1991). Based on the {\it IRAS}
dust trail observations, Sykes \& Walker (1992) found a dust-to-gas ratio of 3.5
for Encke. In fact their dust mass loss loss rate is 
$3\times 10^{13}$ g/orbit, consistent with the result of this paper,
and the gas mass loss rate was taken from an average of historical
visual wavelength observations (Kres\'ak and Kres\'akov\'a 1990).

If Encke continues to lose mass at this rate, its remaining lifetime before
shedding essentially its entire mass is 3000--10,000 $\rho_N$ yr. 
Since Encke is presently in a dynamically
stable orbit, it appears that mass loss will be its cause of death
unless the mass loss is somehow quenched in the next few millenia.
If the mass is lost as a uniform layer over the surface, then the fractional
decrease in surface area of the nucleus per orbit is $2\Delta M/3M$,
which would lead to a reduction in nuclear magnitude by 
0.0002--0.0006 $\rho_N^{-1}$ mag/orbit.
This decrease is significantly less than the average 0.01 mag/orbit secular decrease 
in comet brightnesses inferred by Kres\'ak and Kres\'akov\'a (1990), and it is
far smaller than the aging of 0.09 mag/orbit estimated by Ferrin and Gil (1988).
These conflicts should not be surprising, as the coma brightness is likely a nonlinear function 
of the nuclear surface area, and the derivation of fading from historical observations
is exceptionally difficult. 
As another application of the mass loss rate, we can estimate the decrease in
the nuclear radius with time.
Again assuming the mass is lost as a sheet from the
upper crust of the nucleus, the fractional decrease in nuclear radius per orbit
is $\Delta M/3M$, leading to a change of radius of 0.3-0.9 m/orbit.
If Encke is losing mass like Halley, via jets rather than erosion of the entire
surface layer, then the jets must either be very wide or they must
evolve from place to place on the surface, lest they become so deep that 
sunlight could not illuminate them.

\section{Conclusions}

The ISOCAM image of the near-nucleus region of P/Encke reveals a widespread
region with dust having two principal spatial
components: coma and trail. There is no evidence of a tail composed of 
particles with $\beta >10^{-3}$ (or radius $a<500/\rho$ $\mu$m, with
the density $\rho$ in g~cm$^{-3}$).
The dust structures that are observed within the ISOCAM image require a high
obliquity rotational axis and one or more jets. 
The core of the observed dust trail is dominated by material ejected during 
previous apparitions of comet Encke;  particles from the
current apparition particles have not had enough time
to spread into the trail.

The dust trail near the nucleus (distinguished from the trail well away from
the nucleus as observed by IRAS) appears to have a broad component and a narrow 
component which tracks the comet's orbit. 
The components are `fed' by 
the southern lobe of the coma, which consists mainly of particles with 
$10^{-5} < \beta < 10^{-3}$ (500 $\mu$m--5 cm) and a NW `spike.'
Particles with $\beta = 10^{-4}$ (5 mm) appear to be the delimiter between
broad and narrow trail components, with all particles larger than 5 mm 
($\beta < 10^{-4}$) ending up in the narrow portion of the trail. The broad 
wings of the trail are composed of particles with $10^{-3} < \beta < 10^{-4}$.
This result is consistent with the maximum beta of $\beta \approx 10^{-3}$ 
found in P/Tempel 2 dust trail (Sykes {\it et al.} 1990) suggesting that the 
size distributions of large particles may be similar from comet to comet. 

The dynamical result that P/Encke's dust mass loss is dominated by the emission
of large particles is consistent with the non-detection of a silicate
feature from ISOPHOT observations taken 4 days later (Lisse {\it et al.}
2000). In addition, little or no 10 $\mu$m silicate feature was detected 
during the 1987 apparition, when the inner coma of the comet was observed 
at its brightest, close to perihelion (Gehrz {\it et al.} 1989). The lack of a 
silicate feature, however, does not mean that silicate particles are necessarily
absent. Rather, Gehrz {\it et al.} conclude that the particles are large enough
($a> 10$ $\mu$m) that their emission spectrum does not contain bright 10
$\mu$m (or 20 $\mu$m) spectral features. The broad-band
spectrophotometry both from the 1987 (Gehrz {\it et al.} 1989) and the 1997
(Lisse {\it et al.} 2000) apparitions indicates that the temperature of the
grains in the coma is similar to that expected for rapidly rotating
greybodies. Because the coma
emission spectrum appears similar to that of a greybody out to wavelengths
$>100$ $\mu$m, Lisse {\it et al.} (2000) conclude that the
particles must be larger than  15 $\mu$m in radius. Thus, dynamical and
spectroscopic evidence all point strongly to a
coma dominated by at least mid-sized grains. This distinguishes
Encke from new comets such as C/Hale-Bopp, which evidences fragmentation of
coma dust and generating a significant high-$\beta$ particle population.
The dominance of large particles is even more extreme for
P/Encke than it is for P/Halley (McDonnell et al. 1991; Fulle et al. 1995)
and P/Grigg-Skjellerup (McDonnell et al. 1993),
for which {\it in situ} observations showed the total mass is dominated
by large particles, with mass $>10^{-4}$~g or $a>200 \rho^{-1/3}$ $\mu$m.

Comparing the numerical simulations to the observations, we find the total
mass of particles ejected by Encke in its 1997 apparition is 
2--6 $\times 10^{13}$ g. Most of the particles from the 1997 apparition
are in the extended dust coma around the comet, and
all of these particles eventually end up in the dust trail.
Most of the dust trail in our image consists of particles from previous
apparitions. Comparing the observed surface brightness of the dust trail
to a numerical simulation for 1.5 orbits, and considering the total length
of the dust trail (Sykes and Walker 1992), we infer that the trail is
at least 10 orbits old. Therefore the total mass of the trail is
at least 2--6 $\times 10^{14}$ g. 
Dividing the mass of the nucleus (using $R=2.4$ km [Fernandez et al. 2000]
and a nuclear mass density $\rho_N$ in g~cm$^{-3}$) by the mass production
we infer for the 1997 apparition, the remaining lifetime of comet
Encke is 3000--10,000 $\rho_N$ yr.
The decrease in surface area of the nucleus per orbit is approximately
(0.7-2)/$\rho_N$ \% per century, so it will gradually fade.
Therefore, we are witnessing the final life stages of Encke, before
it eventually disintegrates into a meteoroid stream  or becomes inactive.

\section{Acknowledgments}

We thank Russ Walker for encouraging us to observe comet Encke with
{\it ISO} and for many useful discussions during this project. We
thank the {\it ISO} operations team led by Timo Prusti for getting
the Encke observation scheduled exactly on time. And we thank Mike
A'Hearn for his critical reading of the manuscript and his helpful
comments. This paper benefited from the detailed comments of two anonymoous
referees whom we thank.
MVS acknowledges NASA grant NAG5-3359.

\section{REFERENCES}

\def\pp{\parshape 2 0truecm 16truecm 1truecm 15truecm}
\def\refic #1;#2;#3;#4;#5;#6 {\noindent \pp{\sc #1} {#2}. {#3}. {\it
#4} {\bf #5}, {#6}.}\par 

\def\refpre #1;#2;#3;#4 {\noindent \pp{\sc #1} {#2}. {#3}, {#4}.}\par

\def\refbook #1;#2;#3;#4;#5;#6;#7;#8 {\noindent \pp{\sc #1} {#2}. #3.
In {\it #4} ({#5}, Eds.), pp. #6. #7, #8.} 

\def\refbookpre #1;#2;#3;#4;#5;#6;#7;#8 {\noindent \pp{\sc #1} {#2}.
#3. In {\it #4} ({#5}, Eds.), #6. #7, #8.} 

\refic A'Hearn, M. F., P. V. Birch, P. D. Feldman, and R. L. Millis;1985;Comet
Encke: Gas production and lightcurve;Icarus;64;1--10

\refic Burns, J. A., P. L. Lamy, and S. Soter;1979;Radiation forces
on small particles in the Solar System;Icarus;40;1--48

\refic Campins, H.;1988;The anomalous dust production in periodic
comet Encke;Icarus;73;508--515


\refic Campins, H., R. G. Walker, and D. J. Lien;1990;IRAS images
of comet Tempel 2;Icarus;86;228--235

\refic Ceplecha, Z. K., J. I. Borovicka, W. G. Elford, D. O. Revelle,
R. L. Hawkes, V. Porubcan, and M. Simek;1998;Meteor phenomena and 
bodies;Space Sci. Rev.;84;327--471
		  
\refic Cesarsky, C. J. and 65 colleagues;1996;ISOCAM in flight;Astron.
Astrophys;315;L32--L37

\refic Corrigan, C. M., M. E. Zolensky, J. Dahl, M. Long, K. Weir, C. Sapp, 
and P. J. Burkett;1997;The porosity and permeability of chondritic meteorites and 
interplanetary dust particles;Meteorit.~Planet.~Sci.;32;509 

\refic Coulais, A., and A. Abergel;2000;Transient correction for
the LW-ISOCAM data for low contrasted 
illumination;Astron.~Astrophys.~Suppl;141;533--544

\refic Curdt, W., and H. U. Keller;1990;Large particles along the
Giotto trajectory;Icarus;86;305--313

\refic Davies, J. K., S. F. Green, A. J. Meadows, B. C. Stewart,
and H. H. Aumann;1984;The {\it IRAS} fast-moving object
search;Nature;309;315--319


\refic Fanale, F. P. and J. R. Salvail;1984;An idealized short-period comet 
model, surface insolation, H$_2$O flux, dust flux and mantle 
evolution;Icarus;60;476--511

\refic Fulle, M., L. Colangeli, V. Mennella, A. Rotundi, and E. 
Bussoletti;1995;The sensitivity of the size distribution to the grain dynamics: 
simulation of the dust flux measured by GIOTTO at P/Halley;Astron. 
Astrophys.;304;622--630

\refic Gehrz, R. D., E. P. Ney, J. Piscetelli, E. Rosenthal, and A. T. Tokunaga;
1989;Infrared photometry and spectroscopy of Comet P/Encke 1987;Icarus;80;280--288

\refic Kamoun, P. G., D. B. Campbell, S. J. Ostro, G. H. Pettengill,
and I. I. Shapiro;1982;Comet Encke---Radar detection of
nucleus;Science;216;293--295

\refic Kelsall,  T., J. L. Weiland, B. A. Franz, W. T. Reach,
	R. G. Arendt, E. Dwek, H. T. Freudenreich, M. G. Hauser, 
	S. H. Moseley, N. P. Odegard, R. F. Silverberg, and
	E. L. Wright;1998;The COBE Diffuse Infrared Background Experiment
	search for the cosmic infrared background. II. Model of the 
	interplanetary dust cloud;ApJ;508;44--73
	
\refic Kessler, M. F., and 10 colleagues;1996;The Infrared Space
Observatory (ISO) mission;Astron. Astrophys.;315;L27--L31

\refic Kres\'ak, L., and M. Kres\'akov\'a;1990;Secular brightness decrease of periodic 
comets;Icarus;86;82-92.
		       

\refbook Lien, D.;1992;Numerical simulations of cometary dust;Asteroids, 
Comets, Meteors 1991;A. Harris and E. Bowell;359-362;LPI;Houston

\refic Liou, J. C., and H. A. Zook;1996;Comets as a source of low
eccentricity and low inclination interplanetary dust
particles;Icarus;123;491--502

\refic Lisse, C. M., M. F. A'Hearn, M. G. Hauser, D. J. Lien, 
S. H. Moseley, W. T. Reach, and R. F. Silverberg;1998;Infrared
observations of comets by {\it COBE};Astrophys. J.;496;971--991

\refpre Lisse, C. et al.;2000;Infrared observations of Comet Encke;
	submitted to {\it Icarus}

\refic Luu, J., and D. Jewitt;1990;The nucleus of
P/Encke;Icarus;86;69--81

\refbook McDonnell, J. A. M., P. Lamy, and G. S.
Pankiewicz;1991;Physical properties of cometary dust;Comets in the 
Post-Halley Era;R. L. Newburn, Jr., M. Neugebauer, and 
J. Rahe;1043--1073;Kluwer;Dordrecht

\refic McDonnell, J. A. M., W. M. Alexander, W. M. Burton, E. Bussoletti, D. H. Clark, 
J. L. Grard, E. Gr\"un, M. S. Hanner, Z. Sekanina, and D. W. Hughes;1986;Dust 
density and mass distribution near comet Halley from Giotto 
observations;Nature;321;338--341

\refpre Moneti, A.;1998;Quantitative cross-calibrations: I. Theory
and first results;preprint~(European~Space~Agency)

\refic Newburn, R. L. and H. Spinrad;1985;Spectrophotometry of seventeen
comets. II. The continuum;Astron. J.;90;2591--2608

\refpre Okumura, K.;1998;ISOCAM PSF report;ISO/ESA Memorandum

\refic P\'erault, M., F. X. D\'esert, A. Abergel, F. Boulanger,
Ch. Dupraz, A. Soufflot, C. J. Cesarsky, and
L. G. Vigroux;1994;Infrared Space Observatory camera calibration facility and
preflight characterization;Opt. Eng.;33;762--770

\refbook Reach, W. T., and F. Boulanger;1998;Infrared emission from interstellar
dust in the local interstellar medium;The Local Bubble and 
Beyond;D. Breitschwerdt, M. J. Freyburg and J. 
Tr\"umper;353--362;Springer;Berlin

\refic Reitsema, H. J., W. A. Delamere, A. R. Williams, D. C. Boice, 
W. F. Huebner, and F. L. Whipple;1989;Dust distribution in the inner coma of Comet 
Halley---Comparison with models;Icarus;81;31--40

\refic Schlegel, D. J., D. P. Finkbeiner, and M. Davis;1998;Maps of dust 
infrared emission for use in estimation of reddening and cosmic microwave 
background radiation foregrounds;Astrophys. J.;500;525--553

\refic Sekanina, Z.;1986;Effects of the law for nongravitational forces on the 
precession model of Comet Encke;Astron. J.;91;422--431

\refic Sekanina, Z.;1988a;Outgassing asymmetry of periodic comet Encke. 
I - Apparitions 1924-1984;Astron. J.;95;911--924

\refic Sekanina, Z.;1988b;Outgassing asymmetry of periodic comet Encke. 
II - Apparitions 1868-1918 and a study of the nucleus 
evolution;Astron. J.;96;1455--1475

\refic Sykes, M. V.;1993;Great balls of mire;Nature;363;696--697

\refic Sykes, M. V., L. A. Lebofsky, D. M. Hunten, and F. J.
Low;1986;The discovery of dust trails in the orbits of periodic
comets;Science;232;1115--1117

\refic Sykes, M. V., D. J. Lien, and R. G. Walker;1990;The Tempel 2
dust trail;Icarus;86;236--247

\refic Sykes, M. V., and R. G. Walker;1992;Cometary dust trails. I.
Survey;Icarus;95;180--210


\refpre Veverka, J.;1999;Contour Mission;website {\tt http://www.contour2002.org}

\refic Whipple, F. L.;1950;A  comet model. I. The acceleration of
Comet Encke;Astrophys. J.;111;375--394

\refic Whipple, F. L.;1951;A  comet model. II. Physical relations for 
comets and meteors;Astrophys. J.;113;464--474

\refic Whipple, F. L.;1955;A  comet model. III. The zodiacal 
light;Astrophys. J.;121;750--770

\end{document}